\documentclass[aps,prl,twocolumn,amsmath,amssymb,amsfonts,nofootinbib,long,floatfix]{revtex4}
\usepackage{epsfig,latexsym,bm,epstopdf}

\begin{document}

\title{On the self-similarity of nonhelical magnetohydrodynamic turbulence}

\author{Leonardo Campanelli$^{1}$}
\email{leonardo.campanelli@ba.infn.it}
\affiliation{$^1$Dipartimento di Fisica, Universit\`{a} di Bari, I-70126 Bari, Italy}

\date{\today}


\begin{abstract}
We re-analyze the Olesen arguments on the self-similarity properties of freely evolving,
nonhelical magnetohydrodynamic turbulence. We find that a necessary and sufficient condition
for the kinetic and magnetic energy spectra to evolve self-similarly
is that the initial velocity and magnetic field are not homogeneous functions of space
of different degree, to wit, the initial energy spectra are not
simple powers of the wavenumber with different slopes.
If, instead, they are homogeneous functions of the same degree,
the evolution is self-similar, it proceeds through selective decay,
and the order of homogeneity fixes the exponents of the power laws according to which the
kinetic and magnetic energies and correlation lengths evolve in time.
If just one of them is homogeneous,
the evolution is self-similar and such exponents are completely determined
by the slope of that initial spectrum which is a power law.
The latter evolves through selective decay, while the other spectrum
may eventually experience an inverse transfer of energy.
Finally, if the initial velocity and magnetic field are not homogeneous functions,
the evolution of the energy spectra is still self-similar but, this time, the power-law exponents
of energies and correlation lengths depend on a single free parameter which cannot be determined by scaling arguments.
Also in this case, an inverse transfer of energy may in principle take place during the evolution of the system.
\end{abstract}




\maketitle


{\bf Introduction.} -- The {\it inverse transfer} of energy of nonhelical magnetic fields from small to large scales has been recently
observed in high-resolution numerical integration of unforced magnetohydrodynamic (MHD)
equations~\cite{Zra14,Bra15} (see also~\cite{Ber14}). This new effect in nonhelical turbulence,
which is instead well known to exist in the helical case, may have profound implications in a wide
spectrum of phenomena of physical interest, ranging from gamma ray burst afterglows~\cite{Zra14}
to the evolution of primordial magnetic fields~\cite{Kah15}.
Such an inverse energy transfer has also been found to proceed in a self-similar
way~\cite{Bra15}. Interestingly enough, self-similarity of freely evolving, MHD turbulence was predicted long
time ago by Olesen~\cite{Ole97}.
In this paper, we use the Olesen self-similar solution to the MHD equations to
determine the conditions under which a self-similar evolution, and possibly an inverse energy transfer,
can take place in nonhelical turbulence
(for different interpretations of the Olesen scaling arguments, see~\cite{Ole97,Dit04,You04,Cam04,Ole15,Ol15b}).

{\bf The Olesen solution.} -- It is well known that the MHD equations~\cite{Bis93}, under the scaling transformations
$\textbf{x} \rightarrow \ell \, \textbf{x}$, $t \rightarrow \ell^{1-h} \, t$,
admit a solution of the type~\cite{Ole97}
\begin{equation}
\label{0}
\textbf{z}(\ell \textbf{x},\ell^{1-h} t) = \ell^h \, \textbf{z}(\textbf{x},t),
\end{equation}
where $\ell > 0$ and $h$ are real parameters, and $\textbf{z}$ stands for the velocity of bulk motion,
$\textbf{v}$, or for the magnetic field $\textbf{B}$.
The above ``Olesen solution'', however, is valid provided that the
dissipative parameters $\nu$ (the kinematic viscosity) and $\eta$ (the resistivity) scale as
$\nu(\ell^{1-h} t) = \ell^{1+h} \, \nu(t)$ and $\eta(\ell^{1-h} t) = \ell^{1+h} \, \eta(t)$.
Differentiating the latter equations with respect to $\ell$, and putting $\ell = 1$ afterwards,
we get $\nu(t) \propto \eta(t) \propto t^{(1+h)/(1-h)}$.
As pointed out in~\cite{Ole97}, then, the Olesen solution
is valid for a theory where the dissipation parameters evolve in time as above or, more generally,
in the ``turbulence range'' where dissipation is negligible.
Since we are interested in MHD turbulence, we neglect any dissipation effects and
we work on scales well above the dissipation length. Accordingly, the Olesen solution is valid for all $h$.

The Olesen arguments, however, are useful only for times $t \geq t_* > t_i$, where $t_i$ is the initial time which,
due to translational invariance of MHD equations, we can always take to be zero, $t_i = 0$.
To see this, let us take, for example, $h =-1$, so that the MHD equations are invariant under the re-scaling
$L \rightarrow \ell L $ and $t \rightarrow \ell^2 t$
of lengths $L$ and times $t$. This means that if, for example, we progressively double any physical size $L$ inside the system
taking $\ell = 1,2,4,8,...$, the latter will appear the same if we look at it, respectively, at the times $t_*, 2^2t_*,4^2t_*,8^2t_*,...$.
If we start at $t_i = 0$, instead, we would wait an infinite amount of time in order to see
the system appearing the same after just a single doubling of $L$.
The fact that the Olesen arguments are valid only for $t \geq t_* > t_i$,
can be also physically understood by the fact that turbulence fully develops only after a
certain amount of time has passed from the initial time.

We are interested in the evolution of parity-even,
statistically homogeneous and isotropic MHD turbulence.
This means that the two-point,
divergenceless correlation tensor $C_{ij}({\textbf x},{\textbf y}) = \langle z_i({\textbf x}) z_j({\textbf y}) \rangle$,
where $\langle ... \rangle$ denotes an ensemble average, is a function of
$|{\textbf x}-{\textbf y}|$ only, and it transforms as a polar $SO(3)$ tensor.
Its Fourier transform, $\int \! d^3 x \, e^{i {\textbf k} \cdot
({\textbf x}-{\textbf y})} \, C_{ij}({\textbf x},{\textbf y})$, then, depends only on $k = |{\textbf k}|$.
Accordingly, the kinetic and magnetic energy spectra
$\mathcal{E}_z(k,t) = \frac{k^2}{(2\pi)^2} \! \int \! d^3\!y \, e^{i\textbf{k} \cdot (\textbf{x}-\textbf{y})}
\langle \textbf{z}(\textbf{x},t) \cdot \textbf{z}(\textbf{y},t) \rangle$
are functions of $k$ and $t$.

We analyze the four exhaustive cases: 1) both the kinetic and the magnetic energy spectra are
simple power laws of the wavenumber $k$ with same slope at the initial time $t=t_i$;
2) none of them is a power law; 3) just one of them is a
power law; 4) both of them are power laws, but with different slopes.

{\bf Case 1.} -- Let assume that at the initial time $t_i$,
both the velocity and the magnetic field are self-similar,
$\textbf{z}(\ell \, \textbf{x},t_i) = \ell^h \, \textbf{z}(\textbf{x},t_i)$.
We stress the fact that we are assuming that both $\textbf{v}$ and $\textbf{B}$ scale in the same manner under the
transformation $\textbf{x} \rightarrow \ell \, \textbf{x}$. In mathematical language,
$\textbf{v}(\textbf{x},t_i)$ and $\textbf{B}(\textbf{x},t_i)$ are homogeneous functions of $\textbf{x}$ of the same degree $h$.
This implies that the kinetic and magnetic energy spectra at $t=t_i$ are simple powers of $k$.
In fact, observing that $\mathcal{E}_z(k/\ell,t_i) = \ell^{1+2h} \mathcal{E}_z(k,t_i)$
and defining $\psi_z(k)$ by the relation $\mathcal{E}_z(k,t_i) = k^{-1-2h} \, \psi_z(k)$,
we find $\psi_z(k/\ell) = \psi_z(k)$, which implies that the function $\psi_z(k)$ do not depend on $k$.
Writing $\psi_z(k) = \lambda_z$, with $\lambda_z$ being a constant, we get
$\mathcal{E}_z(k,t_i) = \lambda_z k^{\alpha}$, where we have defined $\alpha = -1-2h$.
Let us assume that $\alpha > -1$, so that we are working under the hypothesis that the degree of homogeneity
of $\textbf{z}(\textbf{x},t_i)$ is negative. In this case,
the kinetic and magnetic energies (see below) at the time $t_i$ are ultraviolet divergent, but infrared finite.
Such an ultraviolet divergence, however, disappears at times $t>t_i$ due to the presence of dissipation
which acts in damping both the velocity and the magnetic fields at very short scales.
To see this, let us make the very rough assumption that from the initial time up to the time when turbulence develops,
the evolution of the system is approximatively regulated by dissipation only.
In this case, we have
$\mathcal{E}_z(k,t) = \mathcal{E}_z(k,0) e^{-k^2 \xi^2_z(t)}$, where
$\xi_v(t) = \sqrt{2\nu t}$ and $\xi_B(t) = \sqrt{2\eta t}$ are
the kinetic and magnetic dissipation lengths (we are assuming, for simplicity, that $\nu$ and $\eta$ are constants).
We see that at the time $t > t_i = 0$, dissipation is acting as an effective cut-off with
cut-off scale(s) equal to the dissipation scale(s).
In particular, if the energy spectra at the initial time are simple power law, then, when turbulence
begins at the time $t = t_* > t_i = 0$, they will be cut-offed power law,
\begin{equation}
\label{101}
\mathcal{E}_z(k,t_*) = \lambda_z k^\alpha e^{-k^2 \xi^2_z(t_*)}.
\end{equation}
Such a kind of dissipative cut-off can also be understood as a
process of ``spreading out'' the velocity and magnetic field
over the dissipation scale. To see this, let us spread out $\textbf{z}(\textbf{x},t_i)$
over $\xi_z(t_*)$ in the following way:
\begin{equation}
\label{1}
\textbf{z}(\textbf{x},t_*) = \int \! d^3\! y \, w(|\textbf{x}-\textbf{y}|/\xi_z(t_*)) \, \textbf{z}(\textbf{y},t_i),
\end{equation}
where $w(|\textbf{x}|/\xi_z(t_*))$ is a window function
normalized to unity, $\int \! d^3\! x \, w(|\textbf{x}|/\xi_z(t_*)) = 1$,
and such that $w(|\textbf{x}|/\xi_z(t_*)) \rightarrow 0$ for $|\textbf{x}|/\xi_z(t_*) \rightarrow \infty$.
Taking into account Eq.~(\ref{1}), the energy spectra at the time $t=t_*$ become
$\mathcal{E}_z(k,t_*) = w^2(k\xi_z(t_*)) \, \mathcal{E}_z(k,t_i)$,
where $w(k\xi_z(t_*))$ is the Fourier transform of $w(|\textbf{x}|/\xi_z(t_*))$.
We note that $w^2(k\xi_z(t_*)) \rightarrow 1$ for $k\xi_z(t_*) \rightarrow 0$, and we will
assume that $w^2(k\xi_z(t_*))$ goes to zero for $k\xi_z(t_*) \rightarrow \infty$
sufficiently fast to ensure finite energies (see below).
Taking a Gaussian window function,
$w(|\textbf{x}|,\xi_z(t_*)) = (2\pi \xi_z^2(t_*))^{-3/2} e^{-|\textbf{x}|^2\!/2\xi_z^2(t_*)}$,
whose Fourier transform is $w(k,\xi_z(t_*)) = e^{-k^2\xi_z^2(t_*) /2}$,
we recover Eq.~(\ref{101}).

Now, we can relax the assumption that from $t_i$ up to $t_*$ the system
evolves just in a dissipative way. What we can do, in the light of the above discussion,
is to average out the effects of dissipation at the time $t_*$ by using Eq.~(\ref{1}),
with $w(|\textbf{x}|/\xi_z(t_*))$ being an unknown window function.
Also, the ``spreading scale'' $\xi_z(t_*)$ is related to the dissipative scale in a way that cannot be determined by scaling arguments.
What we can say is that it must depend on the dissipative parameters $\nu$ and $\eta$ in such a way that
\begin{equation}
\label{lim}
\lim_{\nu,\eta \rightarrow 0} \xi_z(t_*) = 0.
\end{equation}
In this case, then, the energy spectra at the time when turbulence begins will have the form
\begin{equation}
\label{s}
\mathcal{E}_z(k,t_*) = \lambda_z k^{\alpha} w^2(k\xi_z(t_*)).
\end{equation}
We stress here the fact that the above equation is valid
on all scales, since we have taken into account the effects of dissipation by averaging them out.
Scaling arguments alone, however, cannot give the expression of the (dissipative) cut-off
$w^2(k\xi_z(t_*))$, whose exact expression, nevertheless, is inessential for our discussions.


\begin{figure}[t!]
\begin{center}
\includegraphics[clip,width=0.248\textwidth]{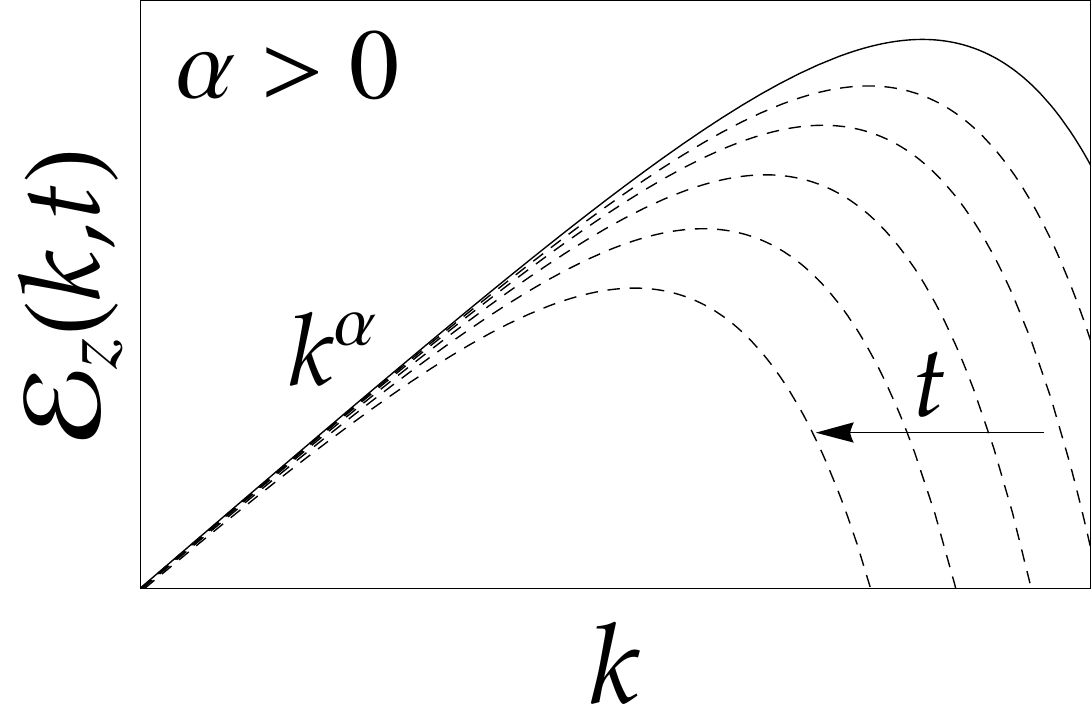}
\hspace{-0.56cm}
\includegraphics[clip,width=0.248\textwidth]{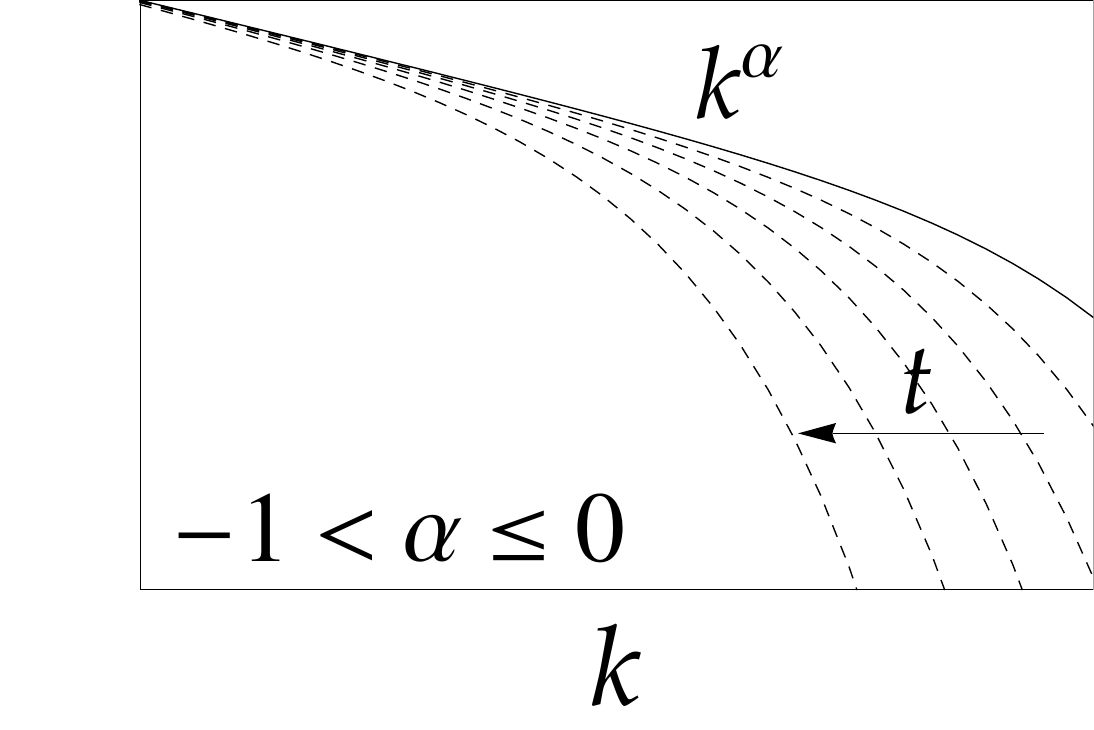}
\caption{Schematic plot (in arbitrary logarithmic units) of the kinetic and/or magnetic energy
spectra $\mathcal{E}_z$ as a function of the wavenumber $k$ as the time (indicated by an arrow)
increases. The spectra at the onset of turbulence at $t=t_*$, represented by continues lines, are 
a cut-off power law of $k$ with slope $\alpha > -1$ 
(cases 1 and 3 in the text). Dashed lines
denote the same spectra at later times. The evolution of turbulence proceeds through selective
decay.}
\end{center}
\end{figure}


Let us now see what the Olesen solution implies for the expression of the kinetic and magnetic energy
spectra at later times. We start by the Olesen result~\cite{Ole97} that, in the
turbulence range, the energy spectra have the form
\begin{equation}
\label{Eqnewnew5}
\mathcal{E}_z(k,t) = k^{\alpha} \psi_z(k^{(3+\alpha)/2} t),
\end{equation}
where $\psi_z$ is an arbitrary (scaling-invariant) function of its argument.
The validity of this result can be extended to all scales if one averages out the
effects of dissipation as discussed above. To do this, we impose the requirement that the spectra
in Eq.~(\ref{Eqnewnew5}) evaluated at $t = t_*$ equal those in Eq.~(\ref{s}).
We find $\psi_z(k^{(3+\alpha)/2} t_*) = \lambda_z w^2(k\xi_z(t_*))$.
Now, this straightforwardly implies that
\begin{equation}
\label{Eqnewnew6}
\psi_z(k^{(3+\alpha)/2} t) = \lambda_z w^2(k\xi_z(t)),
\end{equation}
where we have introduced the time-dependent quantity
\begin{equation}
\label{Eqnewnew7}
\xi_z(t) = \xi_z(t_*) (t/t_*)^{2/(3+\alpha)}.
\end{equation}
We stress that the result in Eq.~(\ref{Eqnewnew6}) is now valid on all scales.
This can be understood as follows. At the time when turbulence begins, we average out the effects of
dissipation by spreading out the velocity and magnetic field over a scale below which dissipation is active.
This ``closes'' the dissipative range and leaves ``open'', in an effective way, just the turbulence range,
$0 \leq k \leq 2\pi/\xi_z(t_*)$.
At later times, the dissipative range remains closed, while the turbulence range, where the Olesen result is valid,
shrinks as $0 \leq k \leq 2\pi/\xi_z(t)$.
In a certain sense, the scaling-invariance of MHD equations in the turbulence range
effectively ``transfers'' the spreading done at $t_*$ at later times. This results in
an effective, time-dependent spreading scale, under which energy is dissipated, which evolves
according to Eq.~(\ref{Eqnewnew7}).

Due to this interpretation, according to which the above expressions of the energy spectra
are valid on all scales, we can compute the kinetic and magnetic energies
$E_z(t) = \frac12 \langle \textbf{z}^2(\textbf{x},t) \rangle = \int_0^{\infty} \!dk \,\mathcal{E}_z(k,t)$.
We find that they decay in time as
\begin{equation}
\label{Eqnew14} E_z(t)  = c_z \, \xi_z(t)^{-(1+\alpha)} \propto t^{-2(1+\alpha)/(3+\alpha)},
\end{equation}
where $c_z = \lambda_z \! \int_0^\infty \! dx w^2(x) x^\alpha$.

Let us observe that the spreading scale $\xi_z(t)$ gives, apart from an inessential numerical factor, the
so-called (kinetic and magnetic) correlation length defined by
$\zeta_z(t) = 2\pi \! \int_0^{\infty} \! dk k^{-1} \mathcal{E}_z(k,t) / \! \int_0^{\infty} \! dk \mathcal{E}_z(k,t)$.
In fact, we have
$\zeta_z(t) = 2\pi \! \left[\int_0^\infty \! dx w^2(x) x^{\alpha-1} \! / \! \int_0^\infty \! dx w^2(x) x^\alpha\right] \!\xi_z(t)$.
In the case $\alpha > 0$, the correlation length is finite and it defines the maximum scale on which the
velocity and magnetic field are correlated.
In the case $-1 < \alpha \leq 0$, instead, $\zeta_z$ is divergent. This implies that
$\textbf{v}$ and $\textbf{B}$ are correlated on arbitrarily large scales. In a real situation,
however, when the system has a finite size, this means that they are
correlated on the maximum allowed scale, i.e., the linear size of the system.

The above result on the form of the kinetic and magnetic energy spectra, pictorially shown in Fig.~1,
implies that the evolution of the system proceeds through ``selective decay'', to wit,
modes with large wavenumber decay faster than those whose wavenumber is small.
The selective-decay evolution of the energy spectra,
as well as the growth
of correlation lengths and the decay of energies, Eqs.~(\ref{Eqnewnew7}) and (\ref{Eqnew14}), respectively,
are in agreement with the analytical
result of~\cite{Cam07} obtained in mean-field approximation.

Let us now briefly compare our results with those obtained in recent high-resolution numerical simulations.
First, let us consider the hydrodynamical case ($\textbf{B} = 0$)
with an initial power-law spectrum $\mathcal{E}_{v}(k,t_i) \propto k^4$,
as considered in~\cite{Bra15}. According to our results, the system should evolve
self-similarly and through selective decay after turbulence develops at $t=t_*$,
with an energy spectrum given by
$\mathcal{E}_{v}(k,t) \propto k^4 w^2(k\xi_v(t))$.
This spectrum can be re-written as
$\mathcal{E}_{v}(k,t) \propto \xi^{-\alpha}_v(t) \phi_v(k \xi_v(t))$, with
$\alpha = 4$ and $\phi_v(x) = x^4 w^2(x)$.
These predictions about self-similarity and selective decay of the system
are in good agreement with the results of~\cite{Bra15}. However, the authors of~\cite{Bra15}
find that the value $\alpha = 3$, instead of $\alpha = 4$, better describes the numerical results.

In the MHD case, let us consider $\mathcal{E}_z(k,t_i) \propto k^\alpha$ with $\alpha > 0$.
The magnetic energy spectrum evolves in time through selective
decay, as observed in~\cite{Ban04}. Moreover, the magnetic energy
decreases in time faster for higher values of $\alpha$, and this is also observed in~\cite{Ban04}.
Finally, the decay law~(\ref{Eqnew14}) is in discrete agreement with the numerical results of~\cite{Ban04}.

{\bf Case 2.} -- Let now assume that both $\textbf{v}$ and $\textbf{B}$ are not
self-similar at the initial time $t_i=0$.
This implies that both the initial kinetic and the initial magnetic energy spectra are not simple power laws of $k$. Proceeding as in the case 1, we find that the energy spectra evolve self-similarly as
\begin{equation}
\label{Olesen} \mathcal{E}_z(k,t) = \xi(t)^{-\alpha} \phi_z(k \xi(t)),
\end{equation}
where $\phi_z(k \xi(t)) = \mathcal{E}_z(k \xi(t),0) \, w^2(k \xi_z(t))$, with $\xi(t) = (t/t_*)^{2/(3+\alpha)}$
and, this time, $\alpha$ is a free parameter.
Also, we easily find that correlation lengths and energies evolve as
$\zeta_z(t) \propto t^{2/(3+\alpha)}$ and $E_z(t) \propto t^{-2(1+\alpha)/(3+\alpha)}$.

Let us suppose that for $k \rightarrow 0$ the initial spectra are
power laws, $\mathcal{E}_z(k \xi(t),0) \propto k^{\alpha_z}$ with exponents $\alpha_z > 0$.
In this case, since $w^2(k \xi_z(t)) \rightarrow 1$ for $k \rightarrow 0$,
we have $\mathcal{E}_z(k,t) \rightarrow k^{\alpha_z} \xi(t)^{\alpha_z-\alpha}$.
Therefore, in this case, an inverse transfer of energy can occur if $\alpha_z > \alpha$ ($\alpha_z < \alpha$)
and $\xi(t)$ is an increasing (decreasing) function of time.

Recently, the scaling relation~(\ref{Olesen}) has been numerically confirmed by
Brandenburg, Kahniashvili, and Tevzadze~\cite{Bra15}. In particular, they numerically found
$\alpha \simeq 1$ and that there is an inverse transfer of energy at large scales.
This can be understood as follows.
Taking, as in~\cite{Bra15}, $\mathcal{E}_v(k,0) \propto k^2$ and $\mathcal{E}_B(k,0) \propto k^4$ for $k \rightarrow 0$,
Eq.~(\ref{Olesen}) with $\alpha = 1$ implies at large scales ($k \rightarrow 0$) an inverse transfer of energy, which goes as
$\mathcal{E}_B(k,t) \propto k^4 t^{3/2}$ for the magnetic field and,
less efficiently, as $\mathcal{E}_{v}(k,t) \propto k^2 t^{1/2}$ for the kinetic case.

{\bf Case 3.} -- Let now assume that at $t = t_i$, just one of the two independent variables
$\textbf{v}$ or $\textbf{B}$ is homogeneous and the other is not. In this case, the exponent $\alpha$ in Eq.~(\ref{Olesen})
is not a free parameter, and it is indeed determined by the slope of the energy spectrum of that variable which is
homogeneous at $t=t_i$. Such a spectrum evolves through selective decay (as in case 1),
while the other may eventually undergo an inverse transfer.

As an example, let us consider a (quasi-) scaling-invariant magnetic field ($\alpha \simeq -1$)
in a cosmological context.
%
%
Such a type of large-scale magnetic field could have been generated
during inflation as a result of the breaking of conformal invariance of standard Maxwell theory~\cite{Tur88,Sub15}.
Generally, an inflation-produced magnetic field is correlated on super-Hubble scales during
inflation, but eventually it re-enters the horizon during radiation- or matter-dominated
eras~\cite{Cam15}. At this time, the magnetic field inside the horizon starts to interact with the
primeval plasma whose bulk velocity is not generally an homogeneous
function of space. The evolution of the kinetic and magnetic field energy spectra
is self-similar in this case, and it is described by Eq.~(\ref{Eqnewnew5}) with $\alpha \simeq -1$.
Causality confines the interaction between the magnetic field and the plasma within the Hubble radius
which represents the physical maximum correlation length at the time of re-entering.
As time passes, the magnetic energy spectrum retains its shape at scales near the Hubble radius
while, at much smaller scales, it decay selectively as pictorially shown in the right panel of Fig.~1.

All these features have been observed in a recent numerical simulation performed in~\cite{Kah12}.
Here, however, an external pumping is at work during the evolution of the magnetic field.
Nevertheless, the energy is injected at very small scales, very far from the
physical maximum correlation length, which coincides with the size
of the computational cube. Hence, it is plausible to believe that
the decay of the magnetic field at large scales is statistically decoupled from the energy source and
then it is just ruled by the self-similarity properties of the (unforced) MHD equations.

{\bf Case 4.} -- Finally, in the case where at the initial time both $\textbf{v}$ and $\textbf{B}$
are homogeneous functions, but with different degrees (which means that the
initial kinetic and magnetic energy spectra are power of $k$ with different slopes),
no self-similar solutions of MHD equations exist in the turbulence regime.

{\bf Helical turbulence.} -- Before concluding, we briefly comment
on some aspects of helical magnetohydrodynamic turbulence. In the presence of parity-odd quantities
like the magnetic helicity, linkages in the vortex lines of the turbulent magnetic flow will be present~\cite{Bis93}.
These linkages, eliminated by resistivity on small scales, survive in the turbulence regime.
Accordingly, the evolution of helical magnetic fields in ideal MHD turbulence
is very different from that of nonhelical fields, as we considered here. In the former case, in fact,
the neglect of dissipation introduces a true topological invariant,
the number of magnetic linkages, which is not present in the latter case.
When considering the helical case, and that is beyond the aim of this paper, any scaling argument
must properly take into account the presence of such a topological invariant.

{\bf Conclusions.} -- Using Olesen scaling arguments, we have found the conditions under which the kinetic and the magnetic energy spectra
evolve self-similarly in freely evolving, nonhelical magnetohydrodynamic turbulence.

If, at the initial time, both the kinetic and the magnetic energy spectra are
power laws of the wavenumber $k$ with same slope,
the evolution is self-similar. In particular, the energy spectra are progressively washed out by selective decay,
while the energies and correlation lengths
evolve as powers of the time, with exponents completely determined by the slope of the
initial energy spectra.

If the initial energy spectra are not power laws,
the evolution is still self-similar but, this time, such exponents
depend on a single parameter which cannot be fixed by scaling arguments.
In this case, an inverse transfer of energy may in principle take place.

If just one of the initial spectra is a power law,
the evolution is self-similar and the exponents of the power laws according to which the
energies and correlation lengths evolve in time are completely determined
by the slope of that initial energy spectrum which is a power law.
Such a spectrum evolves through selective decay, while the other
may eventually go through an inverse transfer.

Finally, if both the initial kinetic and the initial magnetic energy spectra are power laws, but with different slopes,
no self-similar solutions exist.

\vspace{0.1cm}
We would like to thank T.~Kahniashvili for useful discussions.

\end{document}